\documentclass[11pt,letterpaper]{article}

\usepackage[T1]{fontenc}
\usepackage[utf8]{inputenc}
\usepackage{lmodern}
\usepackage{microtype}
\usepackage[margin=1in]{geometry}
\usepackage{booktabs}
\usepackage{tabularx}
\usepackage{longtable}
\usepackage{array}
\usepackage{enumitem}
\usepackage{listings}
\usepackage{xcolor}
\usepackage[hyphens]{url}
\usepackage[numbers]{natbib}
\usepackage[hidelinks,bookmarks=false]{hyperref}

\setlist[itemize]{leftmargin=*,nosep}

\definecolor{codebg}{gray}{0.96}
\newcommand{\dagtoml}{DAG-TOML}
\newcommand{\hellobytes}{\texttt{Hello, world!\textbackslash n}}

\title{\bfseries A Minimal Executable Proof for Multi-Language Contract Traceability}
\author{Werner Kasselman\\Verivus OSS\\\texttt{werner@verivus.com}}
\date{}

\begin{document}
\maketitle

\begin{abstract}
This paper reports a deliberately small executable proof for a
\dagtoml{} contract: six ``Hello, world!'' implementations in Rust, Go,
C, Java, TypeScript, and AWK are linked to one observable-output
contract, one implementation DAG, one traceability file, one readiness
gate, and one evidence matrix. The load-bearing contract requires the
exact UTF-8 byte sequence \hellobytes{}, zero stderr bytes, and exit
code 0. On the runner used for this paper, the witness harness reported
five PASS outcomes, one SKIP for Java because \texttt{javac/java} was
not on \texttt{PATH}, and zero FAIL outcomes. Two sidecar witnesses
exercise narrower source-analysis claims: a convoluted Go rewrite hides
the contiguous greeting literal but remains visible to sqry at the
declared AST symbol and simple-edge level, while an indirect AWK
rewrite uses a declared source profile because AWK is not in the
repository's sqry-backed validator language set. The contribution is
not a benchmark, a claim of general semantic equivalence, or a
production assurance system. It is a compact, falsifiable artifact that
shows how a contract, implementation graph, traceability chain, and
review gate can be checked against executable witnesses.
\end{abstract}

\section{Introduction}

Software-engineering papers about artifacts are easiest to review when
the claims, commands, and files line up directly. Artifact-review
guidelines make the same point operationally: reviewers need enough
structure to connect a paper's claims to executable scripts, source
files, expected outputs, and limitations~\cite{acmBadging2020,poplArtifact2023}.
The \dagtoml{} repositories provide such a structure for engineering
agents and reviewers. This paper examines the smallest nontrivial
instance in the paper artifact repository:
\path{hello-world/proof-bundle}.

The example has one primary runtime contract. Each canonical
implementation must write the exact bytes \hellobytes{} to stdout,
write no bytes to stderr, and exit with code 0. The example also
contains two source-analysis witnesses. The Go witness demonstrates
that a specific rewrite can remove the plain greeting literal while
leaving declared function symbols and simple graph edges visible to
AST-aware tooling. The AWK witness demonstrates a fallback profile for
an unsupported language: because the repository's sqry-backed symbol
validator covers Rust, Go, TypeScript, and Java, not AWK, the AWK
rewrite is checked by a narrower deterministic source-profile script.

This is intentionally a minimal proof, not a performance experiment.
There is no dataset, no timing comparison, no statistical model, and no
claim that all source rewrites can be reduced to author intent. The
goal is to show that the proof pack is inspectable and falsifiable in a
single sitting.

\section{Artifact}

The proof pack lives under \path{hello-world/proof-bundle}. Its
normative files are:

\begin{itemize}
  \item \path{contract_declaration.toml}: contracts C01 through C06.
  \item \path{implementation_dag.toml}: nine units, all tier 1.
  \item \path{traceability.toml}: intent, requirement, implementation,
        code, test, and output chains.
  \item \path{review_readiness.toml}: one gate for the witness pack.
  \item \path{evidence_matrix.toml}: five claims mapped to seven
        evidence artifacts.
  \item \path{run_all.sh}, \path{detect_semantic_rewrite.sh}, and
        \path{detect_awk_rewrite.sh}: executable witnesses.
\end{itemize}

Relative to \path{hello-world/proof-bundle}, the canonical
implementations are \path{src/rust/hello.rs},
\path{src/go/hello.go}, \path{src/c/hello.c},
\path{src/java/Hello.java}, \path{src/typescript/hello.ts}, and
\path{src/awk/hello.awk}. The rewrite fixtures are
\path{src/go_convoluted/hello.go} and
\path{src/awk_convoluted/hello.awk}.

\section{Contract and Witnesses}

Table~\ref{tab:contracts} lists the six declared contracts and the
local witness that supports each claim. C01 is the load-bearing runtime
contract. C02 through C04 are narrow output constraints that depend on
C01. C05 and C06 are source-analysis contracts with explicit
non-claims.

\begin{table}[htbp]
\centering
\small
\begin{tabularx}{\linewidth}{@{}p{0.11\linewidth}p{0.30\linewidth}X@{}}
\toprule
Contract & Domain & Witness and scope \\
\midrule
C01 & Observable output &
\path{run_all.sh}; exact stdout bytes \hellobytes{}, empty stderr, exit
0, with missing toolchains reported as SKIP. \\
C02 & Encoding &
Depends on C01; output must be ASCII bytes that are valid UTF-8 and not
locale-dependent in the tested pipe context. \\
C03 & No terminal markup &
Depends on C01; byte-exact comparison rejects ANSI escapes and other
extra output bytes. \\
C04 & No BOM prefix &
Depends on C01; byte-exact comparison rejects any non-\texttt{H} first
byte, including UTF-8 BOM. \\
C05 & Go AST rewrite detectability &
\path{detect_semantic_rewrite.sh}; literal absence, C01 runtime
satisfaction, declared Go functions, caller edge, and import edge. \\
C06 & AWK source-profile detectability &
\path{detect_awk_rewrite.sh}; literal absence, C01 runtime
satisfaction, shared BEGIN/print intent profile, and rewrite markers. \\
\bottomrule
\end{tabularx}
\caption{Contract-to-witness inventory.}
\label{tab:contracts}
\end{table}

\section{Implementation DAG}

The implementation DAG is a fan-in graph. Six independent layer-0 units
prepare the canonical source entries. A layer-1 verification unit
consumes those source artifacts, then builds or runs only the entries
whose required toolchains are available and enforces C01 on each
executed artifact. Two additional layer-0 units run the independent Go
and AWK rewrite witnesses.

\begin{lstlisting}
U01 rust   \
U02 go      \
U03 c        \
U04 java      > U06 verify-contract-c01
U05 ts       /
U08 awk     /

U07 verify-semantic-ast-rewrite     (independent leaf)
U09 verify-awk-rewrite-detection    (independent leaf)
\end{lstlisting}

The checked computed values are validated by the repository's
implementation-DAG validator:

\begin{itemize}
  \item nine units;
  \item layer counts \texttt{\{0: 8, 1: 1\}};
  \item entry points \texttt{U01, U02, U03, U04, U05, U07, U08, U09};
  \item leaf nodes \texttt{U06, U07, U09};
  \item critical path \texttt{U05 -> U06} with critical-path LOC 138.
\end{itemize}

\section{Observed Execution}

All commands in this section were run from the repository root on
May 27, 2026. The Java toolchain was not available on the runner, so
the Java implementation is not claimed to have executed in this run.
The harness reports that condition as SKIP, not PASS.

\begin{table}[htbp]
\centering
\small
\begin{tabularx}{\linewidth}{@{}p{0.49\linewidth}p{0.11\linewidth}X@{}}
\toprule
Command & Exit & Observed result \\
\midrule
\path{bash hello-world/proof-bundle/run_all.sh} & 0 &
5 pass, 1 skip, 0 fail. Rust, Go, C, TypeScript, and AWK passed; Java
skipped because \texttt{javac/java} was absent. \\
\path{bash hello-world/proof-bundle/detect_semantic_rewrite.sh} & 0 &
8 pass, 0 skip, 0 fail. Go rewrite witness passed. \\
\path{bash hello-world/proof-bundle/detect_awk_rewrite.sh} & 0 &
6 pass, 0 skip, 0 fail. AWK source-profile witness passed. \\
\path{python3 ../agent-assurance/validators/validate_implementation_dag.py ...} & 0 &
Implementation DAG validation passed. \\
\path{python3 ../agent-assurance/validators/validate_traceability.py ... --check-paths-exist} & 0 &
Traceability validation passed with 30 entities and path checks enabled. \\
\path{python3 ../agent-assurance/validators/validate_review_readiness.py ...} & 0 &
Readiness-gate, contract-declaration, and evidence-matrix files passed. \\
\path{python3 ../agent-assurance/validators/validate_ijb_conformance.py <file>} & 1 &
FAIL for each proof TOML when run exactly without \texttt{--repo-root};
the validator requires \texttt{--repo-root} for instance files. \\
\path{python3 ../agent-assurance/validators/validate_ijb_conformance.py <file> --repo-root ../agent-assurance} & 0 &
PASS for all five proof TOML files when the spec repository root is supplied. \\
\path{python3 ../agent-assurance/validators/validate_code_symbols.py ...} & 0 &
8 supported symbols checked and matched; 4 entries skipped for
unsupported languages. \\
\bottomrule
\end{tabularx}
\caption{Observed command outcomes.}
\label{tab:commands}
\end{table}

During preparation, the harness was audited for whether it truly
checked the trailing newline byte. The original shell comparison read
stdout through command substitution, which strips trailing newlines.
The witness scripts were therefore corrected to compare output files
with \texttt{cmp} against an explicit \hellobytes{} byte stream and to
check stderr by file size. The outcomes in Table~\ref{tab:commands}
are from the corrected witnesses.

\section{PASS, SKIP, FAIL, and MEASURED}

The proof uses four result words in a deliberately narrow way.

\begin{itemize}
  \item PASS means the relevant executable witness or validator exited
        0 and reported that the checked condition held.
  \item SKIP means a declared check was not performed because a required
        toolchain was unavailable. SKIP is not evidence that the skipped
        implementation satisfies the contract.
  \item FAIL means the executable witness or validator exited nonzero or
        reported a violated condition. The IJB no-\texttt{--repo-root}
        invocations in Table~\ref{tab:commands} are preserved as FAIL.
  \item MEASURED is reserved for descriptive observations that are not
        pass/fail gates. This proof reports no benchmark or performance
        measurement.
\end{itemize}

\section{Claim Audit}

Table~\ref{tab:claims} maps the paper's claims to the evidence used.
The distinction between direct observation and inference is important:
for example, the source files directly show the implementations, while
the run commands directly show only the implementations whose
toolchains were available.

\begin{longtable}{@{}p{0.24\linewidth}p{0.30\linewidth}p{0.18\linewidth}p{0.20\linewidth}@{}}
\toprule
Claim & Evidence source & Status & Counterexample boundary \\
\midrule
\endhead
The proof pack is structurally complete for this example. &
\path{hello-world/proof-bundle/implementation_dag.toml},
\path{hello-world/proof-bundle/traceability.toml},
\path{hello-world/proof-bundle/contract_declaration.toml},
\path{hello-world/proof-bundle/review_readiness.toml},
\path{hello-world/proof-bundle/evidence_matrix.toml}; validators. &
Directly observed. &
Does not imply production readiness. \\
The available canonical implementations satisfy C01 on this runner. &
\path{hello-world/proof-bundle/run_all.sh} corrected byte comparison and observed output. &
Directly observed for Rust, Go, C, TypeScript, AWK; Java skipped. &
A runner without those toolchains would produce more SKIPs. \\
The Go rewrite hides the literal but exposes declared AST structure. &
\path{hello-world/proof-bundle/src/go_convoluted/hello.go} and
\path{hello-world/proof-bundle/detect_semantic_rewrite.sh}. &
Directly observed. &
Does not prove arbitrary obfuscation resistance. \\
AWK uses a fallback profile because it is unsupported by the sqry symbol
validator. &
the spec repository's
\path{validators/validate_code_symbols.py} and
\path{hello-world/proof-bundle/detect_awk_rewrite.sh}. &
Directly observed plus narrow inference. &
Does not prove broad AWK AST similarity. \\
Artifact organization follows current review expectations. &
ACM badging, POPL artifact guidance, arXiv TeX guidance, and SIGSOFT
standards~\cite{acmBadging2020,poplArtifact2023,arxivSubmitTex2026,sigsoftStandards}. &
Cited. &
No formal artifact badge is claimed. \\
\bottomrule
\caption{Claim-to-evidence audit.}
\label{tab:claims}
\end{longtable}

\section{Related Work}

The proof's source-analysis claims sit in a long line of program
similarity and clone-detection work. Text and token methods are useful
baselines but can be affected by formatting, naming, and local
rewrites; tree-based and graph-based methods inspect program structure
instead~\cite{schleimer2003winnowing,prechelt2002jplag,roy2009clone}.
DECKARD represents programs through AST-derived structural vectors for
scalable clone detection~\cite{jiang2007deckard}, while GumTree is a
well-known AST differencing system for source changes~\cite{falleri2014gumtree}.
Recent robustness work shows why plagiarism-hiding transformations
must be stated carefully and why no single checker should be treated as
a legal or semantic oracle~\cite{cheers2021robustness}.

The artifact framing follows research-artifact guidance rather than
benchmark methodology. ACM distinguishes artifact evaluation,
availability, and result validation~\cite{acmBadging2020}; POPL's
artifact guidance asks authors to map scripts and source files to paper
claims~\cite{poplArtifact2023}; SIGSOFT's empirical standards emphasize
more specific and technical review checklists for software-engineering
research~\cite{sigsoftStandards,ralph2020empirical}. The executable
specification angle also has precedent in trace specifications, where
formal specifications can be turned into executable models used as
consistency checks and prototypes~\cite{hoffman1988trace}.

\section{Threats to Validity}

The validity structure follows common software-engineering reporting
practice for case-study and empirical claims: separate what was
measured from the causal or external claims one might be tempted to
draw~\cite{runeson2009case,feldt2010validity}.

\textbf{Construct validity.} The proof measures whether a tiny contract
is represented and checked, not whether \dagtoml{} is sufficient for a
large assurance pipeline. The Hello-World contract is intentionally
small, so the paper avoids generalizing from it to complex I/O,
sandboxing, supply-chain integrity, or legal provenance.

\textbf{Internal validity.} The strongest internal risk was the shell
newline issue described above. It was corrected before the reported
run. A remaining risk is that the scripts depend on local toolchain
behavior; the Java result is explicitly SKIP on this runner.

\textbf{External validity.} The proof covers six toy implementations
and two hand-written rewrites. It does not show that the same approach
scales to real services, unsafe languages, concurrency, platform-
specific encodings, or adversarial obfuscation.

\textbf{Conclusion validity.} The results are categorical command
outcomes, not statistical estimates. The correct conclusion is that the
declared witnesses passed, skipped, or failed as reported on this
runner. No benchmark, precision, recall, or legal conclusion follows.

\section{arXiv and Artifact Packaging Notes}

The public artifact repository for this paper is
\url{https://github.com/verivus-oss/agent-assurance-papers}. The
specification and validators used by the proof are maintained at
\url{https://github.com/verivus-oss/agent-assurance}.

The paper package is intentionally plain: \texttt{article},
\texttt{pdflatex}, BibTeX, \texttt{natbib}, and standard packages.
Official arXiv guidance says TeX submissions are processed
automatically, authors must inspect the generated PDF, required figures
and bibliography inputs must be included, and extraneous files such as
logs, aux files, backup files, unused figures, and referee material
should be removed from the source package~\cite{arxivSubmitTex2026}.
arXiv currently supports TeX Live 2025 by default and TeX Live 2023 as
a selectable option, with \texttt{pdflatex} among the supported
processors~\cite{arxivTexLive2026}.

For this paper, the artifact is already part of the repository. An
arXiv source package should include only \path{main.tex},
\path{references.bib}, and any generated \path{main.bbl} if submitting
with precomputed BibTeX output. The repository code can be linked from
the paper or uploaded as ancillary material if desired, but the TeX
source package should not include unrelated repository files.

\section{Limitations and Non-Claims}

This paper does not claim:

\begin{itemize}
  \item general semantic equivalence between arbitrary programs;
  \item arbitrary obfuscation resistance;
  \item broad AWK AST similarity or parser-backed AWK analysis;
  \item copyright, licensing, or authorship conclusions;
  \item production readiness for real assurance workflows;
  \item performance, scalability, precision, recall, or benchmark
        results.
\end{itemize}

The exact claim is narrower: the repository contains a small
multi-language proof pack whose declarations and executable witnesses
can be checked, whose unsupported-language boundary is explicit, and
whose observed outcomes are reproducible by running the listed commands
in an environment with the same toolchains.

\section{Conclusion}

The Hello-World proof demonstrates the mechanics of contract-centered
traceability with a small enough artifact that a reviewer can inspect
every moving part. The primary runtime witness passed for five available
toolchains and skipped Java because the required Java tools were absent.
The Go AST witness and AWK source-profile witness passed. The structural
validators passed, and the IJB validator's requirement for
\texttt{--repo-root} was observed rather than hidden. The resulting
artifact is not broad evidence about semantic equivalence or production
assurance. It is a compact, executable proof that claims can be tied to
contracts, DAG nodes, code paths, witnesses, and review gates.

\bibliographystyle{plainnat}
\bibliography{references}

\end{document}